\documentstyle[12pt,aps]{revtex}

\title{Numerical integration of the equations of motion \\
       for rigid polyatomics: The matrix method \\ [6pt]}

\author{\sc Igor P. Omelyan \\ [1.5ex]
{\small \em Institute for Condensed Matter Physics,
            National Ukrainian Academy of Sciences,} \\ [-8pt]
{\small \em 1 Svientsitsky St., UA-290011 Lviv, Ukraine \thanks
           {E-mail: nep@icmp.lviv.ua}} \\}

\newcommand{\bms}[1]{\mbox{\boldmath $#1$}}
\newcommand{\bvs}[1]{\mbox{\scriptsize\boldmath $#1$}}

\begin{document}

\setlength{\abovedisplayskip}{18pt plus2pt minus6pt}
\setlength{\belowdisplayskip}{\abovedisplayskip}
\setlength{\abovedisplayshortskip}{10pt plus2pt minus4pt}
\setlength{\belowdisplayshortskip}{\abovedisplayshortskip}

\maketitle

\vspace{1cm}

\begin{abstract}

A new scheme for numerical integration of motion for classical systems
composed of rigid polyatomic molecules is proposed. The scheme is based
on a matrix representation of the rotational degrees of freedom. The
equations of motion are integrated within the Verlet framework in velocity
form. It is shown that, contrary to previous methods, in the approach
introduced the rigidity of molecules can be conserved automatically
without any additional transformations. A comparison of various
techniques with respect to numerical stability is made.

\vspace{0.75cm}


\vspace{0.1cm}


\end{abstract}

\newpage

\section{Introduction}

\hspace{1em}  Many approaches in physical chemistry deal with model systems
consisting of classical rigid molecules. In studying such systems by the
method of molecular dynamics (MD), three problems arise at least: (a) the
choice of suitable parameters for describing a state of the system in phase
space, (b) the application of an efficient algorithm to integrate numerically
the equations of motion, and (c) the exact conservation of the rigidity of
molecules during an approximate integration. The question of how best to
handle these problems is one which has been keenly debated and the relative
merits of a number of various schemes have been devised.

The molecular approach treats dynamics of the system in view of translational
and rotational motions. In the classical scheme [1], three Eulerian angles
are used to represent the same number of rotational degrees of freedom of
the molecule. A numerical integration of the corresponding equations of
motion was performed in early investigations [2, 3]. It has been soon
established [4, 5] that this integration is very inefficient because of
singularities whenever the azimuthal angle of the molecule takes a value
of $0$ or $\pi$. Although the singularities can be avoided by applying
different sets of Eulerian angles, this requires complex manipulations
with time-consuming trigonometric functions. In singularity free schemes,
the orientations of molecules are expressed in terms of either quaternions
[6--10] or principal-axis vectors [6]. The last scheme has been derived
extending the symmetry vector method [11, 12] for diatomics to an arbitrary
rigid body.

In the atomic approach [13], the phase trajectories are considered as
translational displacements of individual molecular sites. Such particles
move independently under the potential-energy forces and constraint forces,
introduced to hold inter-atomic distances constant. This approach is
intensively exploited in MD simulations since usual algorithms for
integration of translational motion can be applied here. However, the
atomic technique is sophisticated to implement for point molecules and when
there are more than two, three or four interaction sites in the cases of
linear, planar and three-dimensional molecules, respectively, because then
the orientations can not be defined uniquely [14]. Moreover, to reproduce
exactly the rigid molecular structure for arbitrary polyatomics, it is
necessary to solve complicated systems of six nonlinear equations per
molecule at each time step of the integration process.

Usually, high-order predictor-corrector algorithms [15, 16] are applied
to integrate the equations of rotational motion [2, 7, 8]. Such algorithms,
being very accurate at small time steps, quickly become unstable and can
not be used for greater step sizes [14]. Small time steps are impractical
in calculations, because too much expensive computer time is required to
cover the sufficient phase space. At the same time, translational motion
is successfully integrated with the lower-order Verlet [17], leapfrog [18],
velocity Verlet [19] and Beeman [20] algorithms, owing their simplicity
and exceptional numerical stability (for example, the equations of atomic
motion are integrated within the usual Verlet framework [13, 14]). However,
original versions of these algorithms were constructed on an assumption
that acceleration is velocity-independent, and, therefore, they can not
be applied directly to rotational dynamics. Analogous pattern arises with
translational motion in the presence of external magnetic fields or when
relativistic effects are important and it is necessary to take into account
internal fields of moving charges.

To remedy that omission, Fincham [21, 22] has proposed explicit and implicit
versions of the leapfrog algorithm for rotational motion in which angular
momenta are involved into the integration. In the case of a more stable
implicit version, this leads to a system of four nonlinear equations per
molecule for the same number of quaternion components, which is solved by
iteration [22]. Ahlrichs and Brode have derived a hybrid method [23] in
which the principal axes are considered as pseudo particles and constraint
forces are introduced to maintain their orthonormality. The evolution of
principal axes in time can be determined using a recursive solution for
exponential propagators. In such a way some difficulties of the cumbersome
atomic technique have been obviated. But the algorithm is within the Verlet
framework and does not involve angular velocities. Therefore, it is
impossible to extend it to a thermostat version or to an integration in
the presence of magnetic fields. Moreover, the pseudo-particle formalism
does not contain molecular torques, so that it is not so simple matter to
apply it to systems with point multipoles. Finally, the recursive method
[23] as well as the rotational-motion leapfrog algorithms [22] appear to
be much less efficient with respect to the total energy conservation than
the atomic-constraint technique.

In the present paper we develop the idea of using principal-axis vectors as
orientational variables. We involve the velocities and molecular torques
explicitly and show that the rigidity problem can easily be resolved in our
approach without any additional transformations. The paper is organized as
follows. The equations of motion for orientational matrices are obtained in
Sec.2. The question of how to integrate these equations within the Verlet
framework in velocity form is considered in Sec.3. A comparison of different
approaches, based on actual MD simulations of water, is presented in Sec.4.
Concluding remarks are added in Sec.5.

\section{Equations of molecular motion}

\hspace{1em}  Let us consider a system composed of $N$ identical rigid
molecules with $M$ atoms. We split evolution of the system in time $t$
into translational and rotational motions. The translational motions are
applied with respect to the molecule as a whole and can be described by
the $3N$ ($i=1,\ldots,N$) Newton equations
\begin{equation}
m \frac{{\rm d}^2 \bms{r}_i}{{\rm d} t^2} = \mathop{\sum_{j;a,b}^{N;M}}
\limits_{(j \ne i)} \bms{F}_{ij}^{ab} (|\bms{r}_i^a-\bms{r}_j^b|)
\equiv \bms{F}_i(t) \ ,
\end{equation}
where $\bms{r}_i=\sum_a^M m_a \bms{r}_i^a/m$ and $\bms{r}_i^a$ are the
positions of the centre of mass and atom $a$ of molecule $i$, respectively,
$m=\sum_a^M m_a$ and $m_a$ denote the masses of a separate molecule and
partial atoms, and $\bms{F}_{ij}^{ab}$ are the atom-atom forces between
two different molecules.

To analyze rotational motions, we introduce the sets $\bms{e} \equiv
(\bms{e}_1, \bms{e}_2, \bms{e}_3)$ and $\bms{u}^i \equiv (\bms{u}_1^i,
\bms{u}_2^i, \bms{u}_3^i)$ of orthogonal unit vectors characterizing the
laboratory fixed coordinate system, L, and the moving coordinate system,
S$^i$, attached to molecule $i$, respectively. Orientations of the
S$^i$-system with respect to the laboratory frame can be defined as
$\bms{u}_\alpha^i=\sum_\beta a_{\alpha \beta}^i \bms{e}_\beta$, or
merely ${\bms{u}^i}^{\bvs{+}}={\bf A}_i \bms{e}^{\bvs{+}}$, where
$\bms{e}^{\bvs{+}}$ and ${\bms{u}^i}^{\bvs{+}}$ are vector-columns,
$a_{\alpha \beta}^i=\bms{u}_\alpha^i \bms{\cdot} \bms{e}_\beta$ are
components of the rotational matrix ${\bf A}_i$ and $\alpha, \beta =
1,2,3$. Let us place the origin of the S$^i$-system in the centre of
mass of the $i$-th molecule and direct the axes of this system along
the principal axes of inertia. The principal components of angular
velocities, $\bms{\mit \Omega}_i={\mit \Omega}_1^i \bms{u}_1^i+{\mit
\Omega}_2^i \bms{u}_2^i+{\mit \Omega}_3^i \bms{u}_3^i$, obey $3N$
Euler equations [2],
\begin{equation}
J_\alpha \frac{{\rm d} {\mit \Omega}_\alpha^i}{{\rm d} t} =
K_\alpha^i(t) + \Big(J_\beta-J_\gamma\Big) {\mit \Omega}_\beta^i(t)
{\mit \Omega}_\gamma^i(t) \ ,
\end{equation}
where $(\alpha,\beta,\gamma)=(1,2,3)$; $(2,3,1)$ and $(3,1,2)$. Here $J_1$,
$J_2$ and $J_3$ are the independent on time principal moments of inertia of
the molecule, $\sum_{j;a,b} \bms{\delta}_i^a \bms{\times} \bms{F}_{ij}^{ab}
= k_1^i \bms{e}_1+k_2^i \bms{e}_2+k_3^i \bms{e}_3 = K_1^i \bms{u}_1^i+K_2^i
\bms{u}_2^i+K_3^i \bms{u}_3^i$ is the torque exerted on molecule $i$ with
respect to its centre of mass due to the interactions with the other
molecules, $\bms{K}_i^{\bvs{+}}={\bf A}_i \bms{k}_i^{\bvs{+}}$, where
$\bms{K}_i=(K_1^i, K_2^i, K_3^i)$, $\bms{k}_i=(k_1^i, k_2^i, k_3^i)$ and
$\bms{\delta}_i^a=\bms{r}_i^a-\bms{r}_i$. Let $\bms{\Delta}^a=(\Delta^a_1,
\Delta^a_2, \Delta^a_3)^{\bvs{+}}$ be a vector-column of positions for
atom $a$ within the molecule in the S$^i$-system, i.e., $\bms{\delta}_i^a=
\Delta^a_1 \bms{u}_1^i+\Delta^a_2 \bms{u}_2^i+ \Delta^a_3 \bms{u}_3^i$. By
construction of the S$^i$-system the conservative set ($a=1,\ldots,M$) of
vectors $\bms{\Delta}^a$ is the same for each molecule and defined by its
rigid geometry. Then the positions of atoms in the L-system at time $t$ are
$\bms{r}_i^a(t) = \bms{r}_i(t) + {\bf A}_i^{\bvs{+}}(t) \bms{\Delta}^a$,
where ${\bf A}^{\bvs{+}}$ denotes the matrix transposed to ${\bf A}$.

Usually, the elements of orientational matrices ${\bf A}_i$ are expressed
via three Eulerian angles which can be chosen as follows: $\cos \theta_i=
\bms{e}_3 \bms{\cdot} \bms{u}_3^i$, $\cos \varphi_i=\bms{e}_2 \bms{\cdot}
(\bms{e}_3 \bms{\times} \bms{u}_3^i)/|\bms{e}_3 \bms{\times} \bms{u}_3^i|$
and $\cos \psi_i=\bms{u}_2^i \bms{\cdot} (\bms{e}_3 \bms{\times} \bms{u}_3^i)
/|\bms{e}_3 \bms{\times} \bms{u}_3^i|$. Then principal components of angular
velocity are ${\mit \Omega}_1^i=\dot \theta_i \sin \psi_i - \dot \varphi_i
\sin \theta_i \cos \psi_i$, ${\mit \Omega}_2^i= \dot \theta_i \cos \psi_i +
\dot \varphi_i \sin \theta_i \sin \psi_i$ and ${\mit \Omega}_3^i=\dot
\varphi_i \cos \theta_i + \dot \psi_i$. As was mentioned earlier, the
equations of motion are singular in this case. The most notorious
demonstration is the expression $({\mit \Omega}_2^i \sin \psi_i - {\mit
\Omega}_1^i \cos \psi_i)/\sin \theta_i$ for the generalized velocity $\dot
\varphi_i$ from which it follows that $\dot \varphi_i \to \infty$ when
$\theta_i$ tends to zero or $\pi$. This leads to serious technical
disadvantages for the application of Eulerian angles to numerical
calculations. It is worth mentioning that the rigidity of molecules is
conserved automatically in this approach, i.e., $|\bms{\delta}_i^a(t)|^2=
\left({\bf A}_i^{\bvs{+}}(t) \bms{\Delta}^a \right)^{\bvs{+}} \left({\bf
A}_i^{\bvs{+}}(t) \bms{\Delta}^a \right)={\bms{\Delta}^a}^{\bvs{+}} {\bf
A}_i(t) {\bf A}_i^{\bvs{+}}(t) \bms{\Delta}^a =|\bms{\Delta}^a|^2$, where
the property ${\bf A} {\bf A}^{\bvs{+}}={\bf I}$ of rotational matrices
has been used and ${\bf I}$ is the unit matrix. In other words, the matrix
${\bf A}_i$ remains an orthonormal one for arbitrary values of Eulerian
angles.

As is now well established [6, 7], at least four orientational parameters
per molecule must be used to avoid the singularities. In this case the
matrix ${\bf A}_i$ is a function of these parameters which constitute the
so-called quaternion $\bms{q}_i \equiv (\xi_i, \eta_i, \zeta_i, \chi_i)$.
It is necessary to emphasize that the matrix ${\bf A}_i$ is orthonormal
if the quaternion satisfies the equality $\bms{q}_i^2=\xi_i^2+\eta_i^2+
\zeta_i^2+\chi_i^2=1$. In practice, however, the equations of motion are
not solved exactly, so that this constraint will only be satisfied
approximately. The simplest way to achieve the required unit norm at all
times lies in multiplying each quaternion component, associated with the
same molecule, by the common factor $1 \Big/ \sqrt{\bms{q}_i^2}$ at every
time step of the numerical integration [7, 22].

In the mentioned above approaches, orientations of the S$^i$-system with
respect to the laboratory frame were defined by the matrix ${\bf A}_i$,
where each of the nine elements $|a_{\alpha \beta}^i| \le 1$ of which is
a function of either three Eulerian angles or four quaternion components.
Now involving no Eulerian angles and quaternions, we merely consider all
these elements as parameters which represent the rotational degrees of
freedom. The elements $a_{\alpha \beta}^i$ are, in fact, the Cartesian
coordinates of principal axes of molecules in the laboratory frame. They
are not independent as it follows from the requirement ${\bf A}_i {\bf
A}_i^{\bvs{+}}={\bf I}$ imposed on rotational matrices. For example,
the first three elements $a_{11}^i$, $a_{12}^i$ and $a_{13}^i$ can be
expressed via the rest of others from the vector relation $\bms{u}_1^i=
\bms{u}_2^i \bms{\times} \bms{u}_3^i$, reducing the number of orientational
parameters per molecule from $9$ to $6$. The remaining six elements
are connected by three constraints, namely, $\bms{u}_2^i \bms{\cdot}
\bms{u}_2^i=1$, $\bms{u}_3^i \bms{\cdot} \bms{u}_3^i=1$ and $\bms{u}_2^i
\bms{\cdot} \bms{u}_3^i=0$. Thus, among these six elements we can choose
arbitrarily three ones, not belonging the same row of the matrix, to form
an independent set, but only with a few exceptions. Indeed, let $a_{21}^i$,
$a_{22}^i$ and $a_{33}^i$ be chosen as independent elements and one
considers a particular case, when $a_{33}^i=\pm 1$. Then from the equality
$\bms{u}_3^i \bms{\cdot} \bms{u}_3^i=1$ it immediately follows that
$a_{31}^i=a_{32}^i=0$. From the next equality $\bms{u}_2^i \bms{\cdot}
\bms{u}_3^i=0$ we find that $a_{23}^i=0$ and, finally, the third relation
$\bms{u}_2^i \bms{\cdot} \bms{u}_2^i=1$ yields the constraint ${a_{21}^i}^
{\!\!\!2}+{a_{22}^i}^{\!\!\!2}$=1 concerning the variables which were
assumed to be as independent quantities.

The reason of this situation is similar to that existing in the case of using
Eulerian angles, where the singularities have appeared at $\theta_i=0$ or
$\pi$, i.e., at $a_{33}^i=\cos \theta_i = \pm 1$. It indicates again about
the impossibility to derive singularity free equations of motion involving
only three orientational variables per molecule. As was pointed out earlier,
four orientational parameters avoid the singularities for arbitrary
polyatomic molecules. Nevertheless, a larger number of parameters can also
be acceptable, but this leads to an increased number of constraints. For
instance, there is one constraint per molecule for quaternion variables,
while there are three or even six constraints for six or nine parameters in
our case. From this point of view, such an original presentation [6] of the
matrix approach has no advantages with respect to the quaternion method.
However, as we shall show in the next section using some specific properties
of the matrix representation, the constraints can be satisfied intrinsically
within particular integration frameworks without any additional
transformations.

The equations of motion for dynamical variables $a_{\alpha \beta}^i$ can be
found as follows. From the definition ${\rm d} \bms{u}_\alpha^i /{\rm d} t
= \bms{\mit \Omega}_i \bms{\times} \bms{u}_\alpha^i$ of angular velocity
and the orthonormality of sets $\bms{e}$ and $\bms{u}^i$, we obtain
$\dot a_{1 \alpha}^i = {\mit \Omega}_3^i a_{2 \alpha}^i - {\mit \Omega}_2^i
a_{3 \alpha}^i$, $\dot a_{2 \alpha}^i = {\mit \Omega}_1^i a_{3 \alpha}^i -
{\mit \Omega}_3^i a_{1 \alpha}^i$ and $\dot a_{3 \alpha}^i = {\mit
\Omega}_2^i a_{1 \alpha}^i - {\mit \Omega}_1^i a_{2 \alpha}^i$, or
in the matrix form
\begin{equation}
{\bf \dot A}_i=\bms{\Omega}_i {\bf A}_i \ ,
\end{equation}
where
\begin{equation}
\bms{\Omega}_i = \left(
\begin{array}{ccc}
0 & {\mit \Omega}_3^i & -{\mit \Omega}_2^i \\
-{\mit \Omega}_3^i & 0 & {\mit \Omega}_1^i \\
{\mit \Omega}_2^i & -{\mit \Omega}_1^i & 0
\end{array}
\right) \\ [4pt]
\end{equation}
are antisymmetric matrices associated with angular velocities $\bms{\mit
\Omega}_i$. Then differentiating relations (3) with respect to time,
one obtains the $9N$ ($i=1,\ldots,N$) scalar equations of motion
\begin{equation}
{\bf \ddot A}_i=\bms{\dot{\Omega}}_i {\bf A}_i +
\bms{\Omega} {\bf \dot A}_i = \bms{\dot{\Omega}}_i {\bf A}_i
+ \bms{\Omega}_i \bms{\Omega}_i {\bf A}_i \ ,
\end{equation}
where $\bms{\dot \Omega}_i$ are defined according to Euler equations
(2) and angular velocities are excluded from equalities (3), i.e.,
$\bms{\Omega}_i={\bf \dot A}_i {\bf A}_i^{\bvs{+}}$. In such a way we
construct the coupled set (1), (5) of $12N$ differential equations
of type ${\cal F}(\{ \bms{r}_i, \bms{\ddot r}_i, {\bf A}_i, {\bf \dot
A}_i, {\bf \ddot A}_i\})=0$ in terms of the $12N$ generalized
coordinates $\{\bms{r}_i, {\bf A}_i\}$. If an initial state $\{\bms
{r}_i(t_0), \bms{\dot r}_i(t_0), {\bf A}_i(t_0), {\bf \dot A}_i
(t_0)\}$ is specified, the time evolution $\{\bms{r}_i(t), {\bf A}_i
(t)\}$ of the system can be unambiguously determined by (1) and (5).

\section{Integration within the velocity Verlet framework}

\hspace{1em}  The equations of motion obtained must be complemented by
an integration algorithm in order to be applicable for actual simulations.
As was demonstrated for the atomic approach [13, 14], a very efficient
technique follows from the Verlet algorithm. The same framework has been
used in the pseudo-particle formalism [23]. However, the Verlet algorithm
in its original form [17] does not involve velocity explicitly into the
integration process and, therefore, it can not be applied to equations of
motion with velocity-dependent accelerations, as in our case (see eq.~(5)).
Because of this we shall work within a velocity form [19] of the Verlet
method.

\subsection{Basic ideas}

\hspace{1em}  Let $\{\bms{r}_i(t_0), \bms{\dot r}_i(t_0), {\bf A}_i(t_0),
{\bf \dot A}_i(t_0)\}$ be a spatially-velocity configuration of the system
at time $t_0$. On the basis of equations (1) for translational motion we
can calculate the translational accelerations $\bms{\ddot r}_i(t_0)$ using
molecular forces $\bms{F}_i(t_0)$. Then, according to the first line of
the velocity Verlet integrator, the positions of the centres of mass of
molecules ($i=1,\ldots,N$) at time $t_0+{\mit\Delta}t$ are
\begin{equation}
\bms{r}_i(t_0+{\mit\Delta}t) = \bms{r}_i(t_0) + \bms{\dot r}_i(t_0)
{\mit\Delta}t + \bms{\ddot r}_i(t_0) {\mit\Delta}t^2/2 +
{\cal O}({\mit\Delta}t^3) \ ,
\end{equation}
where ${\mit\Delta}t$ is the time step. Analogously, basing on the equations
for rotational motion (2), we define angular accelerations $\bms{\dot{\mit
\Omega}}_i$ and, therefore, two-fold time derivatives ${\bf \ddot A}_i(t_0)$
(5), using principal torques $\bms{K}_i(t_0)$ and taking into account that
$\bms{\Omega}_i={\bf \dot A}_i {\bf A}_i^{\bvs{+}}$. So that the matrices
${\bf A}_i$ at time $t_0+{\mit \Delta}t$ can be evaluated as follows
\begin{equation}
{\bf A}_i(t_0+{\mit\Delta}t) = {\bf A}_i(t_0) +
{\bf \dot A}_i(t_0) {\mit\Delta}t + {\bf \ddot A}_i(t_0)
{\mit\Delta}t^2/2 + {\cal O}({\mit\Delta}t^3) \ .
\end{equation}

And now we consider how to perform the second line
\begin{equation}
\dot s(t_0+{\mit\Delta}t) = \dot s(t_0) + \Big( \ddot s(t_0) +
\ddot s(t_0+{\mit\Delta}t) \Big) {\mit\Delta}t /2 +
{\cal O}({\mit\Delta}t^3)
\end{equation}
of the velocity Verlet framework, where $s$ denotes a spatial coordinate.
There are no problems to pass this step in the case of translational motion,
when $\dot s \equiv \bms{\dot r}_i$ and, therefore, for new translational
velocities one obtains
\begin{equation}
\bms{\dot r}_i(t_0+{\mit\Delta}t) = \bms{\dot r}_i(t_0) +
\Big( \bms{\ddot r}_i(t_0) + \bms{\ddot r}_i(t_0+{\mit\Delta}t)
\Big) {\mit\Delta}t/2 + {\cal O}({\mit\Delta}t^3) \ ,
\end{equation}
where $\bms{\ddot r}_i(t_0+{\mit\Delta}t)=\frac1m \bms{F}_i(t_0+{\mit
\Delta}t)$ and the forces $\bms{F}_i(t_0+{\mit\Delta}t)$ are calculated
in the already defined new spatial configuration $\{\bms{r}_i(t_0+{\mit
\Delta}t), \bms{\rm A}_i(t_0+{\mit\Delta}t)\}$.

However, the difficulties immediately appear in the case of rotational
dynamics, because then second time derivatives $\ddot s$ can depend
explicitly not only on spatial coordinates $s$, associated with the
rotational degrees of freedom, but also on generalized velocities $\dot
s$. For example, according to Euler equations (2), the principal angular
accelerations depend on orientational variables via molecular torques
and on angular velocities of molecules as well. Then, choosing $s \equiv
{\bf A}_i$, we obtain on the basis of the equations of motion (5) that
${\bf \ddot A}_i(t) \equiv {\bf \ddot A}_i({\bf A}_i(t), {\bf \dot A}_i(t))$.
In view of (8), this leads to a very complicated system of nine nonlinear
equations per molecule with respect to the nine unknown elements of matrix
${\bf \dot A}_i(t+{\mit\Delta}t)$. It is worth to note that similar problems
arise within the leapfrog and Beeman frameworks (see Appendix, where a
rotational-motion version of the Beeman algorithm is derived).

An alternative has been found in rotational-motion versions [21, 22] of the
leapfrog algorithm. It has been assumed to associate the quantity $\dot s$
with the angular momentum $\bms{l}_i = {\bf A}_i^{\bvs{+}} \bms{L}_i$ of
the molecule in the laboratory system of coordinates, i.e., $\dot s \equiv
\bms{l}_i$, where $\bms{L}_i = (J_1 {\mit \Omega}_1^i, J_2 {\mit \Omega}_2^i,
J_3 {\mit \Omega}_3^i) = {\bf J} \bms{{\mit \Omega}}_i$ and ${\bf J}$ is the
diagonal matrix of principal moments of inertia. The rate of change in time
of angular momentum is the torque, i.e., ${\bms{\dot l}}_i=\bms{k}_i$. Then
equation (8) leads to a much more simple expression,
\begin{equation}
\bms{l}_i(t_0+{\mit\Delta}t) = \bms{l}_i(t_0) +
\Big( \bms{k}_i(t_0) + \bms{k}_i(t_0+{\mit\Delta}t) \Big) {\mit\Delta}t /2 +
{\cal O}({\mit\Delta}t^3) \ ,
\end{equation}
and, therefore, new angular momenta are easily evaluated using the known
torques $\bms{k}_i$ at times $t_0$ and $t_0+{\mit\Delta}t$. The corresponding
values for principal angular velocities and first time derivatives of
orientational matrices can be obtained, when they are needed, using the
relations $\bms{\mit \Omega}_i(t_0+{\mit\Delta}t)={\bf J}^{-1} {\bf A}_i(t_0+
{\mit\Delta}t) \bms{l}_i(t_0+{\mit\Delta}t)$ and ${\bf \dot A}_i(t_0+{\mit
\Delta}t) = \bms{\Omega}_i(t_0+{\mit\Delta}t) {\bf A}_i(t_0+{\mit\Delta}t)$.

Finally, we consider the third version of the velocity Verlet method. The
idea consists in using angular velocities as independent parameters for
describing the sate of the system in phase space. Then putting $\dot s
\equiv \bms{\mit \Omega}_i$ in (8) and taking into account Euler
equations (2), we find the following result
\vspace{4pt}
\begin{eqnarray}
{\mit\Delta} {\mit \Omega}_\alpha^i = \frac{{\mit\Delta}t}{2 J_\alpha}
\bigg[ K_\alpha^i(t) + K_\alpha^i(t_0+{\mit\Delta}t) +
\Big(J_\beta-J_\gamma\Big) \hspace{3.5cm} \nonumber \\ [-9pt] \\ [-9pt]
\times \Big( 2 {\mit \Omega}_\beta^i(t_0) {\mit \Omega}_\gamma^i(t_0) +
{\mit \Omega}_\beta^i(t_0) {\mit\Delta} {\mit \Omega}_\gamma^i+
{\mit \Omega}_\gamma^i(t_0) {\mit\Delta} {\mit \Omega}_\beta^i+
{\mit\Delta} {\mit \Omega}_\beta^i {\mit\Delta} {\mit \Omega}_\gamma^i
\Big) \bigg] \ . \hspace{-0.8mm}
\nonumber
\end{eqnarray}
The equations (11) constitute the system of maximum three nonlinear
equations per molecule with respect to the same number of the unknowns
${\mit\Delta} {\mit \Omega}_\alpha^i={\mit \Omega}_\alpha^i(t_0+{\mit
\Delta}t)-{\mit \Omega}_\alpha^i(t_0)$. The system (11) can be linearized,
substituting initially ${\mit\Delta} {\mit \Omega}_\alpha^i=0$ in all
quadratic terms, and solved in a quite efficient way by iteration. This
is justified for ${\mit\Delta}t \to 0$ because then terms nonlinear
in ${\mit\Delta} {\mit \Omega}_\alpha^i$ are small.

From the mathematical point of view, all the three representations $\dot s
\equiv {\bf \dot A}_i, \bms{l}_i$ or $\bms{\mit \Omega}_i$ are completely
equivalent, because the knowledge of an arbitrary quantity from the set
$({\bf \dot A}_i, \bms{l}_i, \bms{\mit \Omega}_i)$ allows us to determine
the rest of two ones uniquely. In the case of numerical integration the
pattern is different, because then the investigated quantities are evaluated
approximately. The choice $\dot s \equiv {\bf \dot A}_i$ can not be
recommended for calculations due to its complexity. The case of $\dot s
\equiv \bms{l}_i$, corresponding to the angular-momentum version of the
Verlet algorithm, is the most attractive in view of the avoidance of
nonlinear equations. Actual computations show, however, that the best
numerical stability with respect to the total energy conservation is
reached in the angular-velocity version (11) of the Verlet algorithm,
when $\dot s \equiv \bms{\mit \Omega}_i$. This fact can be explained
taking into account that a kinetic part, $\frac12 \sum_{i=1}^N (J_1 {{\mit
\Omega}_1^i}^2+J_2 {{\mit \Omega}_2^i}^2+J_3 {{\mit \Omega}_3^i}^2)$, of
the total energy is calculated directly from principal angular velocities.
At the same time, to evaluate angular velocities within the angular-momentum
version the additional transformations $\bms{\mit \Omega}_i={\bf J}^{-1}
{\bf A}_i \bms{l}_i$ with approximately computed matrices ${\bf A}_i$ and
angular momenta $\bms{l}_i$ are necessary. They contribute additional
portions into the accumulated errors at calculations of the total energy.

Shifting the initial time $t_0$ to $t_0+{\mit\Delta}t$, the integration
process is repeated for a next time step. In such a way, step by step
the dynamics of the system can be evaluated.

\subsection{Solving the rigidity problem}

\hspace{1em}  Let us write an analytical solution for orientational
matrices in the form
\begin{equation}
{\bf A}_i(t_0+{\mit\Delta}t)=\sum_{p=0}^P {\bf A}_i^{(p)}(t_0)
\frac{{\mit\Delta}t^p}{p!} + {\cal O}({\mit\Delta}t^{P+1}) \ ,
\end{equation}
where ${\bf A}_i^{(p)}(t_0)$ denotes the $p\,$-fold time derivative of
${\bf A}_i$ at time $t_0$. It can be shown easily from the structure of
equation (3) that arbitrary-order time derivatives of the matrix constraint
$\bms{\Theta}_i(t) \equiv {\bf A}_i(t) {\bf A}_i^{\bvs{+}}(t)-\bms{\rm I}=0$
are equal to zero at a given moment of time, i.e., ${\bf \dot A}_i {\bf
A}_i^{\bvs{+}}+{\bf A}_i {\bf \dot A}_i^{\bvs{+}}=0$, ${\bf \ddot A}_i {\bf
A}_i^{\bvs{+}}+ 2 {\bf \dot A}_i {\bf \dot A}_i^{\bvs{+}}+{\bf A}_i {\bf
\ddot A}_i^{\bvs{+}}=0$ and so on, when ${\bf A}_i$ is orthonormal.
Therefore, if all the terms $(P \to \infty)$ of Taylor's expansion (12)
are taken into account, that corresponds to the exact solution of equations
of motion, and initially all the constraints are satisfied, $\bms{\Theta}_i
(t_0)=0$, they will be fulfilled at later times as well.

In particular algorithms the expansion is truncated after a finite number of
terms. For example, the velocity-Verlet form (7) is restricted by quadratic
terms ($P=2$), involving truncation errors of order ${\mit\Delta}t^3$ into
the matrix elements of ${\bf A}_i$. The same order of uncertainties will be
accumulated in $\bms{\Theta}_i(t)$ at each time step, breaking the molecular
structure, i.e., $\bms{\Theta}_i(t_0+{\mit\Delta}t)={\cal O}({\mit\Delta}
t^3)$. In such a case the molecules collapse and can even be destroyed
completely after a sufficient period of time. Therefore, the problem arises:
how to modify the first line of the algorithm to achieve the exact rigidity
for arbitrary times?

\subsubsection{Constraint-matrix scheme}

\hspace{1em}  The usual way to reduce orientational matrices to orthonormal
form lies in using the constraint technique. The main idea is simple. As
far as the elements of orientational matrices are not independent, this
requires, generally speaking, the necessity of introducing additional forces
which appear as a result of the constraints $\bms{\Theta}_i(t)=0$. These
matrix constraints constitute, in fact, six independent scalar relations
per molecule, namely,
\begin{eqnarray}
&&\phi_1^i \equiv
{a_{11}^i}^{\!\!\!2}+{a_{12}^i}^{\!\!\!2}+{a_{13}^i}^{\!\!\!2}-1=0 \ , \ \ \
\phi_4^i \equiv
a_{11}^i a_{21}^i+a_{12}^i a_{22}^i+a_{13}^i a_{23}^i=0 \ , \nonumber \\
&&\phi_2^i \equiv
{a_{21}^i}^{\!\!\!2}+{a_{22}^i}^{\!\!\!2}+{a_{23}^i}^{\!\!\!2}-1=0 \ , \ \ \
\phi_5^i \equiv
a_{11}^i a_{31}^i+a_{12}^i a_{32}^i+a_{13}^i a_{33}^i=0 \ , \\
&&\phi_3^i \equiv
{a_{31}^i}^{\!\!\!2}+{a_{32}^i}^{\!\!\!2}+{a_{33}^i}^{\!\!\!2}-1=0 \ , \ \ \
\phi_6^i \equiv
a_{21}^i a_{31}^i+a_{22}^i a_{32}^i+a_{23}^i a_{33}^i=0 \ . \nonumber
\end{eqnarray}
Then the corresponding constraint forces, acting on dynamical variables
$a_{\alpha \beta}^i$, are $G_{\alpha \beta}^i=-\sum_{l=1}^6 \lambda_l^i
\partial \phi_l^i/\partial a_{\alpha \beta}^i$ or in the matrix
representation
\begin{equation}
{\bf G}_i=-\bms{\Lambda}_i {\bf A}_i \equiv -\left(
\begin{array}{ccc}
2 \lambda_1^i & \lambda_4^i   & \lambda_5^i   \\
\lambda_4^i   & 2 \lambda_2^i & \lambda_6^i   \\
\lambda_5^i   & \lambda_6^i   & 2 \lambda_3^i
\end{array}
\right) {\bf A}_i \ ,
\end{equation}
where $\bms{\Lambda}_i$ are symmetric matrices of Lagrange multipliers.
The matrices of constraint forces are now added in the equations of motion
(5) and, as a consequence, the evaluation of matrix elements (7) is modified
to
\begin{equation}
{\bf A}_i(t_0+{\mit\Delta}t) = {\bf A}_i(t_0) +
{\bf \dot A}_i(t_0) {\mit\Delta}t + {\bf \ddot A}_i(t_0)
{\mit\Delta}t^2/2 + {\bf G}_i(t_0) {\mit\Delta}t^2/2 +
{\cal O}({\mit\Delta}t^3) \ .
\end{equation}

In view of equations (14) and (15), to satisfy the conditions
$\bms{\Theta}_i(t_0+{\mit\Delta}t)=0$ it is necessary to solve the
system $\phi_l^i(t_0+{\mit\Delta}t)=0$ of six ($l=1,\ldots,6$) nonlinear
equations per molecule for six unknown Lagrange multipliers $\lambda_l^i
(t_0)$. As usually, such a system is linearized and the unknowns are
found by iteration. The iteration procedure can be initiated by putting
$\lambda_l^i=0$ in all nonlinear terms and the iterations converge rapidly
at actual step sizes to the physical solutions $\lambda_l^i(t_0) \sim
{\mit\Delta}t$. The contributions of constraint forces into the matrix
evaluation (15) are of order ${\mit\Delta}t^3$, i.e., the same order as
truncation errors of the basic algorithm (7), but the rigidity is now
fulfilled perfectly for arbitrary times in future. It is worth to remark
that the constraint forces introduced should be treated as pseudo forces,
because they depend on details of the numerical integration in a
characteristic way and disappear if the equations of motion are
solved exactly, i.e., when ${\mit\Delta}t \to 0$.

\subsubsection{Rotational-matrix scheme}

\hspace{1em}  Fortunately, the cumbersome procedure of solving nonlinear
equations to preserve the molecular rigidity can be avoided in our approach
using the fact that actual algorithms are accurate to a finite order only
in time step. In view of equalities (3) and (5), the evaluation (7) can be
presented in a more compact form,
\begin{equation}
{\bf A}_i (t_0+{\mit\Delta}t) = {\bf D}_i (t_0,{\mit\Delta}t)
{\bf A}_i (t_0) + {\cal O}({\mit\Delta}t^3) \ ,
\end{equation}
where
\begin{equation}
{\bf D}_i (t_0,{\mit\Delta}t) = {\bf I} + \bms{\Omega}_i(t_0)
{\mit\Delta}t + \left(\bms{\dot {\Omega}}_i(t_0) + \bms{\Omega}_i^2(t_0)
\right) {\mit\Delta}t^2/2
\end{equation}
are evolution matrices. Let the rigidity has been satisfied at time $t_0$,
i.e., $\bms{\Theta}_i(t_0) = {\bf A}_i(t_0) {\bf A}_i^{\bvs{+}}(t_0)-{\bf
I}=0$. Then $\bms{\Theta}_i(t_0+{\mit\Delta} t)={\bf D}_i(t_0,{\mit\Delta}t)
{\bf D}_i^{\bvs{+}}(t_0,{\mit\Delta}t)-{\bf I}={\cal O}({\mit\Delta}t^3)$
or, in other words, the matrices ${\bf D}_i$ are not orthonormal.

The simplest way to present the evolution matrices and, as a consequence,
the orientational matrices in orthonormal form lies in the following. Taking
into account that $\bms{\Omega}_i^2={\bf W}(\bms{\mit \Omega}_i) - {\mit
\Omega_i}^2 {\bf I}$, where ${\mit \Omega_i}=\sqrt{{{\mit \Omega}_1^i}^2+
{{\mit \Omega}_2^i}^2+{{\mit \Omega}_3^i}^2}$ is the magnitude of the
angular velocity and ${\bf W}(\bms{\mit \Omega}_i)$ is a symmetric matrix
with the elements ${\mit \Omega}_\alpha^i {\mit \Omega}_\beta^i$, we rewrite
(17) as
\begin{equation}
{\bf D}_i (t_0,{\mit\Delta}t) = (1- {\mit \Omega_i}^2(t_0)
{\mit\Delta}t^2/2) {\bf I} + {\bf W}_i (\bms{\mit \Omega}_i(t_0))
{\mit\Delta}t^2/2 + \overline{\bms{\Omega}}_i(t_0) {\mit\Delta}t \ ,
\end{equation}
where $\overline{\bms{\Omega}}_i(t_0)$ is an antisymmetric matrix of type
(4), constructed on the mean value $\overline{\bms{\mit \Omega}}_i(t_0)=
\bms{\mit \Omega}_i(t_0)+\bms{\dot {\mit \Omega}}_i(t_0) {\mit\Delta}t/2$
of the angular velocity for the $i$-th molecule during the time interval
$[t_0, t_0 + {\mit\Delta}]$. It is easy to see that replacing $\bms{\mit
\Omega}_i$ by $\overline{\bms{\mit \Omega}}_i$ in (18), we introduce the
error of order ${\mit\Delta}t^3$. Moreover, taking into account that
\begin{equation}
{\mit\Delta}t = \frac{\sin(\overline{\mit \Omega}_i {\mit\Delta}t)}
{\overline{\mit \Omega}_i} + {\cal O}({\mit\Delta}t^3) \ , \ \ \ \ \
\frac{{\mit\Delta}t^2}{2} = 1 - \frac{ \cos(\overline{\mit \Omega}_i
{\mit\Delta}t)}{\overline{\mit \Omega}_i^2} + {\cal O}({\mit\Delta}t^4) \ ,
\end{equation}
we adjust (18) to the form
\begin{equation}
\bms{\cal D}_i(t_0,{\mit\Delta}t) = {\bf I} \cos (\overline{\mit
\Omega}_i {\mit\Delta}t) + \frac{1-\cos (\overline{\mit \Omega}_i
{\mit\Delta}t)}{\overline {\mit \Omega}_i^2} {\bf W}(\overline{\bms
{\mit \Omega}}_i) + \frac{\sin(\overline{\mit \Omega}_i {\mit\Delta}t)}
{\overline{\mit \Omega}_i} \overline{\bms{\Omega}}_i \equiv
\exp(\overline{\bms{\Omega}}_i(t_0) {\mit\Delta}t) \ .
\end{equation}

Let us expand the matrix $\bms{\cal D}_i(t_0,{\mit \Delta}t)$ into the
Taylor's series with respect to ${\mit\Delta}t$. Then it can be verified
easily that each elements of this matrix coincides with the corresponding
element of ${\bf D}_i(t_0,{\mit\Delta}t)$ (17) up to the second order in
${\mit\Delta}t$ inclusively. Higher order terms, being associated with
time derivatives of angular accelerations, are not taken into account
within the velocity Verlet framework and they can merely be omitted
without loss of the precision. Therefore, the matrices $\bms{\cal D}_i
(t_0,{\mit \Delta}t)$ (20) and ${\bf D}_i(t_0,{\mit\Delta}t)$ (17) differ
between themselves by terms of order ${\mit\Delta}t^3$ or higher that is
completely in the self-consistency with truncation errors of the algorithm
considered. However, the main advantage of using $\bms{\cal D}_i$, instead
of ${\bf D}_i$, in the evaluation
\begin{equation}
{\bf A}_i (t_0+{\mit\Delta}t) = \bms{\cal D}_i (t_0,{\mit\Delta}t)
{\bf A}_i (t_0) + {\cal O}({\mit\Delta}t^3) \ ,
\end{equation}
of orientational variables consists in the fact that the matrix $\bms{\cal
D}_i(t_0,{\mit \Delta}t)$ is orthonormal, i.e., $\bms{\cal D}_i(t_0,{\mit
\Delta}t) \bms{\cal D}_i^{\bvs{+}}(t_0,{\mit\Delta}t)={\bf I}$ and then
$\bms{\Theta}_i(t_0+{\mit\Delta}t)=\bms{\cal D}_i(t_0,{\mit\Delta}t)
\bms{\cal D}_i^{\bvs{+}}(t_0,{\mit\Delta}t)-{\bf I}=0$. As it follows from
the structure of eq.~(20), the matrix $\bms{\cal D}_i(t_0,{\mit \Delta}t)$
defines the three-dimensional rotation on angle $\overline{\mit \Omega}_i
{\mit\Delta}t$ around the axis directed along vector $\overline{\bms{\mit
\Omega}}_i$. In such a way, the rigid structures of molecules can be
reproduced exactly at each time step of the integration.

\section{Numerical tests and discussion}

\hspace{1em}  We now test our matrix method on the basis of simulations on
a TIP4P model [24] of water. This method was used by us previously [25]
investigating a Stockmayer fluid of point dipoles. In the TIP4P model
the water molecule consists of four sites, $M=4$. We have used a system
of $N=256$ molecules and the interaction site reaction field geometry [26].
Intersite components of the TIP4P potential were cut off and shifted to zero
at point of the truncation to avoid the system energy drift associated with
the passage of the sites through the surface of the cut-off sphere. The
cut-off radius was half the basic cell length. The MD simulations were
performed in the microcanonical ensemble at a density of 1 g/cm$^3$ and
at a temperature of 298 K. The numerical stability of solutions to the
equations of motion was identified in terms of relative fluctuations,
${\cal E}(t)=\sqrt{\langle (E-\langle E \rangle_t)^2 \rangle_t /\langle E
\rangle_t^2}$, of the total energy $E$ of the system during time $t$.

We have made a comparative test carrying out explicit MD runs using our
angular-velocity Verlet integrator (eq.~(11)) within constraint- and
rotational-matrix schemes (eqs.~(15) and (21), respectively), as well as
the implicit quaternion leapfrog algorithm [22], the pseudo-particle
formalism [23] and the atomic-constraint technique [13]. The runs were
started from an identical well equilibrated configuration. All the
algorithms required almost the same computer time per step (96\% being
spent to evaluate pair interactions). For the purpose of comparison the
quaternion integration with the Gear predictor-corrector algorithm of fifth
order [15, 16] has been considered as well. At least two corrector steps
were used to provide an optimal performance of the predictor-corrector
scheme and, as a consequence, twice or more larger computer time was taken
in this case than that is normally necessary.

The results obtained for relative total energy fluctuations as functions
of the length of the simulations at four fixed step sizes, ${\mit \Delta}t=$
1, 2, 3 and 4 fs, are presented in fig.~1 (water is most commonly simulated
with a step size of order 2 fs [27]). At small time steps, ${\mit \Delta}t
\le$ 1 fs, all the approaches exhibited similar equivalence in the energy
conservation (subset (a) of fig.~1), except the Gear algorithm which produced
much more accurate trajectories. But the Gear algorithm begins to be unstable
already at ${\mit \Delta}t =$ 1 fs and leads to the worst results for ${\mit
\Delta}t \ge$ 1.5 fs (see, as an example, the case ${\mit \Delta}t=$ 2 fs,
subset (b)). Somewhat better stability is observed in the leapfrog and
pseudo-particle approaches. However, at moderate and great time steps,
${\mit \Delta}t \ge$ 2 fs (figs.~1 (b)--(d)), the results are rather poor,
especially in the case of the leapfrog scheme. The best numerical stability
has been achieved with the atomic-constraint algorithm and our matrix method,
which conserve the total energy approximately with the same accuracy up to
${\mit \Delta}t =$ 3 fs. It can be seen easily that the matrix method works
better within the rotational-matrix scheme, so that there is no need to use
the complicated constraint-matrix procedure. Quite a few iterations (the mean
number of iterations per molecule varied from 3 to 5 at ${\mit \Delta}t=
1 \div 4$ fs) was sufficient to obtain solutions to the system of quadratic
equations (11) with a relative iteration precision of $10^{-12}$. This
contributes a negligible small portion additionally into the total
computer time.

To demonstrate that the exact reproduction of molecular rigidity is so
important, we have also integrated the equations of motion in a situation
(eq.~(7)) when no additional normalization and orthogonalization of
principal-axis vectors are used. In this case the total energy fluctuations
increased drastically with increasing the length of the runs at arbitrary
time steps (see, for instance, the corresponding curve in fig.~1 (a)). The
same words can be said in the case when no quaternion renormalization is
applied along the leapfrog trajectories. This is so because in the
free-normalization regime, the structure of molecules is broken that leads
to an unpredictable discrepancy in the calculation of potential forces and
significant deviations of the total energy. We have also established that
the numerical stability is very sensitive to the way of how the quaternion
renormalization is performed. In particular, the energy conservation can be
somewhat improved if the quaternions are renormalized inside the iterative
loop of the implicit leapfrog integrator rather than at the end of each time
step only, as was originally proposed [22].

No shift of the total energy has been observed for the atomic-constraint
and matrix approaches over a length of 10 ps at ${\mit \Delta}t \le$ 3 fs.
Instead, it oscillates around a stable value of $E_0=-33.6$ kJ/mol. To
reproduce the features of microcanonical ensembles quantitatively, it
is necessary for the ratio $\Gamma={\cal E}/\Upsilon$ of total energy
fluctuations to fluctuations $\Upsilon$ of the potential energy to be
no more than a few per cent. For the system under consideration $\Upsilon
\approx 0.56 \%$, so that, for example, the level ${\cal E} = 0.03 \%$ will
correspond to $\Gamma \sim 5\%$ that is still acceptable for precision
calculations. The ratios $\Gamma$, obtained within various approaches at
the end of 10 ps runs, are plotted in fig.~2 as dependent on the time
increment. The results of fig.~2 show that a level of $\Gamma = 5\%$ is
achieved at the time steps 1.2, 1.4, 3.0 and 4.0 fs within the leapfrog,
pseudo-particle, matrix and atomic approaches, respectively. Therefore,
the last two methods allow a step size more than twice larger than the
pseudo-particle and leapfrog algorithms. The functions $\Gamma({\mit
\Delta}t)$ can be interpolated with a great accuracy as $C {\mit \Delta}
t^2+C' {\mit \Delta}t^3$ with the coefficients $C \approx$ 0.28 and 0.30
\% fs$^{-2}$, $C' \approx$ 0.01 and 0.10 \% fs$^{-3}$ for the atomic and
matrix approaches, respectively. The characteristic square growth of
$\Gamma$ at small time steps is completely in line with ${\cal O}({\mit
\Delta}t^2)$ order of global errors for the algorithm considered.

It is worth to underline that analyzing the system over a significantly
shorter time period, of order 1 ps say (as was done by Ahlrichs and Brode
[23]), one may come to a very misleading idea about the energy conservation.
We can see clearly from fig.~1 that such a simulation period (corresponding
to 1000, 500, 333 and 250 time steps at ${\mit \Delta}t=$ 1, 2, 3 and 4 fs,
respectively) is quite insufficient to give a realistic pattern on global
errors accumulated in the total energy. And only beginning from lengths of
order 10 ps, we are entitled to formulate true conclusions on the numerical
stability. These lengths are sufficiently long to observe an appreciable
modification of the system. For instance, during 10 ps even long-lived
dipole moment correlations vanish completely [28]. Moreover, the phase
trajectories of 10 ps long are also sufficient, as a rule, to reproduce
thermodynamic, structure and other properties of water with a reliable
statistical accuracy. The investigation of some collective effects, such
as dielectric relaxation, may require extremely long simulations (up to
1000 ps [28]) to reduce statistical noise. As a result, even the best
algorithms may not provide a required numerical stability. In such a
situation, we can merely slightly rescale the velocities of particles
when the total energy has exceeded an allowed level. Obviously, the
investigated quantities will be little affected by this rescaling if
it is applied not more frequently than after a period of time during
which the correlations have significantly decayed.

In view of the results obtained in this section, we can conclude that
the method proposed appears to be the most efficient among all known
algorithms deriving within the molecular framework and can be considered
as a good alternative to the cumbersome atomic technique. The fact that
our molecular Verlet algorithm conserves the total energy at great step
sizes somewhat worse than the atomic Verlet algorithm results from the
introduction of velocities. As far as velocities appear explicitly, the
angular accelerations begin to be velocity dependent. Further, the angular
velocities are calculated with one step errors of order ${\cal O}({\mit
\Delta}t^3)$ and the same order of uncertainties will be presented
simultaneously in angular accelerations. This, in its turn (see eq.~(11)),
leads to additional terms of orders ${\cal O}({\mit \Delta}t^4)$ and ${\cal
O}({\mit \Delta}t^3)$ in the truncation and global errors, respectively,
for angular velocities and, as a consequence, for the total energy. That
is why in the case of rotational motion the coefficient $C'$ corresponding
to the velocity Verlet differs significantly from that obtained for the
usual (free of velocities) Verlet algorithm. At the same time, the
corresponding values of $C$ are practically equal between themselves, and,
therefore, we may stay about the equivalence of the both algorithms with
respect to the main term of global errors.

The pointed out above minor disadvantage is compensated, however, by a
much more major advantage of our method with respect to the atomic scheme
in that the velocity Verlet algorithm allows to perform simulations in
canonical ensembles. As is well known [22], thermostat calculations can
be carried out with significantly greater step sizes than those used in
microcanonical ensembles. A thermostat version of the velocity Verlet
algorithm for rotational motion will be studied in a separate
investigation.

\section{Conclusion}

\hspace{1em}  We have shown that the difficulties in numerical integration
of rigid polyatomics can be overcame using an alternative approach. In our
singularity free scheme, orientational matrices were used to represent the
rotational degrees of freedom of the system. Although this introduces extra
equations per molecule and the lack of independence for the matrix elements,
but presents no numerical difficulties. An elegant procedure, built directly
into the Verlet algorithm, has allowed to perfectly fulfil the rigidity of
molecules at each step of the trajectory without any additional efforts and
loss of precision. Avoidance of the necessity to solve complex nonlinear
equations for preservation of the molecular rigidity should be a benefit
of the matrix method with respect to the atomic-constraint approach.

We have demonstrated on the basis of actual calculations that the matrix
method leads to results comparable in efficiency with the cumbersome
atomic-constraint technique. The advantages of the matrix scheme are
that it can be implemented for arbitrary rigid bodies, extended to a
thermostat version and realized in MD programmes in a more simple way.

\vspace{12pt}

{\bf Acknowledgements.} The author would like to acknowledge financial
support of the President of Ukraine and to thank D. Fincham for a useful
discussion.

\newpage
\small

\begin{center}
{\large \bf Appendix}
\end{center}
\setcounter{equation}{0}
\renewcommand{\theequation}{A\arabic{equation}}

We now consider the question of how to adopt our matrix scheme to integrate
the equations of motion within the Beeman framework. According to the usual
Beeman algorithm [20], the translational positions and velocities of
molecules are evaluated as
\begin{eqnarray}
&&\bms{r}_i(t_0+{\mit\Delta}t) = \bms{r}_i(t_0) + \bms{\dot r}_i(t_0)
{\mit\Delta}t + [{\textstyle \frac23} \bms{\ddot r}_i(t_0) -
{\textstyle \frac16} \bms{\ddot r}_i(t_0-{\mit\Delta}t)] {\mit\Delta}t^2 +
{\cal O}({\mit\Delta}t^4) \ , \nonumber \\ [-12pt] \\ [-12pt]
&&\bms{\dot r}_i(t_0+{\mit\Delta}t) = \bms{\dot r}_i(t_0) +
[{\textstyle \frac13} \bms{\ddot r}_i(t_0+{\mit\Delta}t) +
{\textstyle \frac56} \bms{\ddot r}_i(t_0) -
{\textstyle \frac16} \bms{\ddot r}_i(t_0-{\mit\Delta}t)]
{\mit\Delta}t + {\cal O}({\mit\Delta}t^3) \ . \nonumber \ \ \ \ \
\end{eqnarray}
The order of truncation errors in coordinates increases to four because
the expression $[{\textstyle \frac23} \bms{\ddot r}_i(t_0) - {\textstyle
\frac16} \bms{\ddot r}_i(t_0-{\mit\Delta}t)] {\mit\Delta}t^2$ can be
reduced to the form $\bms{\ddot r}_i(t_0) {\mit\Delta}t^2/2 + \bms{\dot
{\ddot r}}_i(t_0) {\mit\Delta}t^3/6 + {\cal O}({\mit\Delta}t^4)$ with the
estimation $\bms{\dot{\ddot r}}_i(t_0)=[\bms{\ddot r}_i(t_0)-\bms{\ddot
r}_i(t_0-{\mit\Delta}t)]/{\mit\Delta}t + {\cal O}({\mit\Delta}t)$ of
superaccelerations. The fractions in the second line of eq. (A1) are
obtained in such a way to provide the third order of truncation errors
in velocities and to satisfy exactly the St$\ddot{\rm o}$rmer central
difference approximation [16, 29] of accelerations
\begin{equation}
s(t_0+{\mit\Delta}t) = -s(t_0-{\mit\Delta}t) + 2 s(t_0) +
\ddot s(t_0) {\mit\Delta}t^2 + {\cal O}({\mit\Delta}t^4)
\end{equation}
with $s \equiv \bms{r}_i$. Acting in the spirit of the Beeman framework,
we can write analogous to (A1) equations for orientational matrices and
angular velocities. The result is
\begin{eqnarray}
{\bf A}_i(t_0+{\mit\Delta}t) \!\!\!&=&\!\!\! {\bf A}_i(t_0) +
{\bf \dot A}_i(t_0) {\mit\Delta}t +
[{\textstyle \frac23} {\bf \ddot A}_i(t_0)
- {\textstyle \frac16} {\bf \ddot A}_i(t_0-{\mit\Delta}t)] {\mit\Delta}t^2
\nonumber \\ [-12pt] \\ [-12pt]
\!\!\!&-&\!\!\! [{\textstyle \frac23} \bms{\Lambda}_i(t_0) {\bf A}_i(t_0)
- {\textstyle \frac16} \bms{\Lambda}_i(t_0-{\mit\Delta}t)
{\bf A}_i(t_0-{\mit\Delta}t)] {\mit\Delta}t^2
+ {\cal O}({\mit\Delta}t^4) \ , \nonumber
\end{eqnarray}
\begin{eqnarray}
{{\mit \Omega}_\alpha^i}^{(n+1)}(t_0+{\mit\Delta}t) =
{\mit \Omega}_\alpha^i(t_0) + \frac{{\mit\Delta}t}{J_\alpha}
\bigg[ {\textstyle \frac13} K_\alpha^i(t_0+{\mit\Delta}t)
+ {\textstyle \frac56} K_\alpha^i(t) -
{\textstyle \frac16} K_\alpha^i(t_0-{\mit\Delta}t)
+ \Big(J_\beta-J_\gamma\Big) \nonumber \\ [-9pt] \\ [-9pt]
\times \Big( {\textstyle \frac13}
{{\mit \Omega}_\beta^i}^{(n)}(t_0+{\mit\Delta}t)
{{\mit \Omega}_\gamma^i}^{(n)}(t_0+{\mit\Delta}t) +
{\textstyle \frac56} {\mit \Omega}_\beta^i(t_0)
{\mit \Omega}_\gamma^i(t_0) - {\textstyle \frac16}
{\mit \Omega}_\beta^i(t_0-{\mit\Delta}t)
{\mit \Omega}_\gamma^i(t_0-{\mit\Delta}t) \Big) \bigg] \ , \nonumber
\end{eqnarray}
where the symmetric constraint matrices $\bms{\Lambda}_i(t_0) \sim {\mit
\Delta}t^2$ are found from the constraint relations ${\bf A}_i(t_0+{\mit
\Delta}t) {\bf A}_i^+(t_0+{\mit\Delta}t)={\bf I}$, whereas new values
${\mit \Omega}_\alpha^i(t_0+{\mit\Delta}t)$ for principal components of
the angular velocities can be computed by iteration $(n=0,1,\ldots)$
taking ${{\mit \Omega}_\alpha^i}^{(0)}(t_0+{\mit\Delta}t)={\mit
\Omega}_\alpha^i(t_0)$ as initial guesses.

A rotational-matrix scheme can be derived within the Beeman method as
follows. Consider first a more general procedure for the orthonormalization
of orientational matrices, which will be valid for integrators of arbitrary
order in truncation errors. Let the algorithm applied uses Taylor's
expansion (12) for the time evaluation of orientational matrices.
Then the evolution matrices can be cast as
\begin{equation}
{\bf D}_i (t_0,{\mit\Delta}t) = {\bf I} + \sum_{p=1}^P {\bf D}_i^{(p)}(t_0)
\frac{{\mit\Delta}t^p}{p!} \ ,
\end{equation}
where ${\bf D}_i^{(p)}={\bf A}_i^{(p)} {\bf A}_i^+$, or more explicitly:
${\bf D}_i^{(1)}=\bms{\Omega}_i$, ${\bf D}_i^{(2)}=\bms{\dot {\Omega}}_i+
\bms{\Omega}_i^2$, ${\bf D}_i^{(3)}=\bms{\ddot {\Omega}}_i+2 \bms{\dot
{\Omega}}_i \bms{\Omega}_i+\bms{\Omega}_i \bms{\dot {\Omega}}_i+\bms
{\Omega}_i^3$, ${\bf D}_i^{(4)}=\bms{\dot{\ddot {\Omega}}}_i+3 \bms{\ddot
{\Omega}}_i \bms{\Omega}_i+\bms{\Omega}_i \bms{\ddot {\Omega}}_i+3 \bms{\dot
{\Omega}}_i \bms{\dot {\Omega}}_i+3 \bms{\dot {\Omega}}_i \bms{\Omega}_i^2
+2 \bms{\Omega}_i \bms{\dot {\Omega}}_i \bms{\Omega}_i+\bms{\Omega}_i^2
\bms{\dot {\Omega}}_i+\bms{\Omega}_i^4$ and so on. A rotational-matrix
counterpart of (A5) we find in the orthonormal form
\begin{equation}
\textstyle \bms{\cal D}_i (t_0,{\mit\Delta}t) = \exp \Big(
\sum_{p=1}^P {\bf H}_i^{(p)}(t_0) \frac{{\mit\Delta}t^p}{p!} \Big) \ ,
\end{equation}
where ${\bf H}_i^{(p)}$ are unknown antisymmetric matrices, i.e., ${{\bf
H}_i^{(p)}}^+=-{\bf H}_i^{(p)}$, and expand the exponent (A6) into the
Taylor series at ${\mit\Delta}t \to 0$. It is obvious that $\bms{D}_i
(t_0,{\mit\Delta}t)$ and $\bms{\cal D}_i (t_0,{\mit\Delta}t)$ will be
identical at $P \to \infty$, if all their matrix coefficients,
corresponding to the same powers $p=1,2,\ldots,P$ of ${\mit\Delta}t$,
are equal between themselves. This condition leads to a recursive procedure
with the solutions ${\bf H}_i^{(1)}=\bms{\Omega}_i$, ${\bf H}_i^{(2)}=
\bms{\dot {\Omega}}_i$, ${\bf H}_i^{(3)}=\bms{\ddot {\Omega}}_i+\frac12
(\bms{\dot {\Omega}}_i \bms{\Omega}_i-\bms{\Omega}_i \bms{\dot {\Omega}}_i)$,
${\bf H}_i^{(4)}=\bms{\dot{\ddot {\Omega}}}_i+\bms{\ddot {\Omega}}_i
\bms{\Omega}_i-\bms{\Omega}_i \bms{\ddot {\Omega}}_i$ and so on. The Beeman
approach is accurate to third order in coordinates ($P=3$), i.e.,
$[{\textstyle \frac23} {\bf \ddot A}_i(t_0) - {\textstyle \frac16} {\bf
\ddot A}_i(t_0-{\mit\Delta}t)] {\mit\Delta}t^2={\bf \ddot A}_i(t_0)
{\mit\Delta}t^2/2 + {\bf \dot{\ddot A}}_i(t_0) {\mit\Delta}t^3/6 +
{\cal O}({\mit\Delta}t^4)$, where the superaccelerations
${\bf \dot{\ddot A}}_i(t_0)=[{\bf \ddot A}_i(t_0)-{\bf \ddot A}_i(t_0-
{\mit\Delta}t)]/{\mit\Delta}t+{\cal O}({\mit\Delta}t)$. Similarly we can
estimate angular superaccelerations, $\bms{\ddot {\Omega}}_i(t_0)=[\bms{\dot
{\Omega}}_i(t_0)-\bms{\dot {\Omega}}_i(t_0)(t_0-{\mit\Delta}t)]/{\mit
\Delta}t$, and obtain in this case
\begin{equation}
\bms{\cal D}_i^{\rm B}(t_0,{\mit\Delta}t) = \exp\!\Big( \bms{\Omega}_i(t_0)
{\mit\Delta}t \!+\! [{\textstyle\frac23} \bms{\dot {\Omega}}_i(t_0)\!-\!
{\textstyle\frac16} \bms{\dot {\Omega}}_i(t_0\!-\!{\mit\Delta}t)] 
{\mit\Delta}t^2\!+\![\bms{\dot {\Omega}}_i(t_0) \bms{\Omega}_i(t_0)\!-\!
\bms{\Omega}_i(t_0) \bms{\dot {\Omega}}_i(t_0)]
{\textstyle\frac{{\mit\Delta}t^3 }{12}} \Big) \ .
\end{equation}
Putting $P=2$ in eq.~(A6) yields the result $\bms{\cal D}_i(t_0,{\mit
\Delta}t) = \exp( \bms{\Omega}_i(t_0) {\mit\Delta}t + {\textstyle\frac12}
\bms{\dot {\Omega}}_i(t_0) {\mit\Delta}t^2)$. As was expected, this is
completely in line with the result (20) performed in Sec.3 for the
velocity Verlet algorithm on the basis of intuitive grounds.

It is worth mentioning that approximation (A2) is used directly for
evaluation of spatial coordinates in the usual Verlet algorithm [17, 23]. As
can be verified, any trajectory produced by the velocity Verlet algorithm
satisfies equation (A2) at $s \equiv \{\bms{r}_i, {\bf A}_i\}$ even if
constraint- or rotational-matrix schemes are used. The fact that the
trajectory $s(t)$ can be generated with the same fourth-order local
errors by lower-order equations (6) and (7) (or (21)) results from a
fortunate cancellation of truncation errors arising in coordinates and
velocities during two neighbour time steps. Note, however, that the usual
Verlet algorithm, its velocity version and Beeman method are not equivalent,
because they differ between themselves by the main term of fourth-order
uncertainties in coordinates and calculate one-step velocities in a
different manner. For example, evaluating velocities within the Beeman
algorithm, it is assumed that the accelerations are slow variables on time
scales of $2{\mit\Delta}t$. If this criterion is not satisfied, the main
term ${\cal O}({\mit\Delta}t^3)$ of truncation uncertainties in velocities
and, as a result, the main coefficient $C$ in global errors for the total
energy may increase in a characteristic way. This prediction has been
confirmed by our computer simulations on the TIP4P water. Therefore, the
Beeman algorithm can be applied for systems with sufficiently smooth
interparticle potentials only.

\newpage
\begin{center}

\vspace*{1cm}

{\large Figure captions}
\end{center}

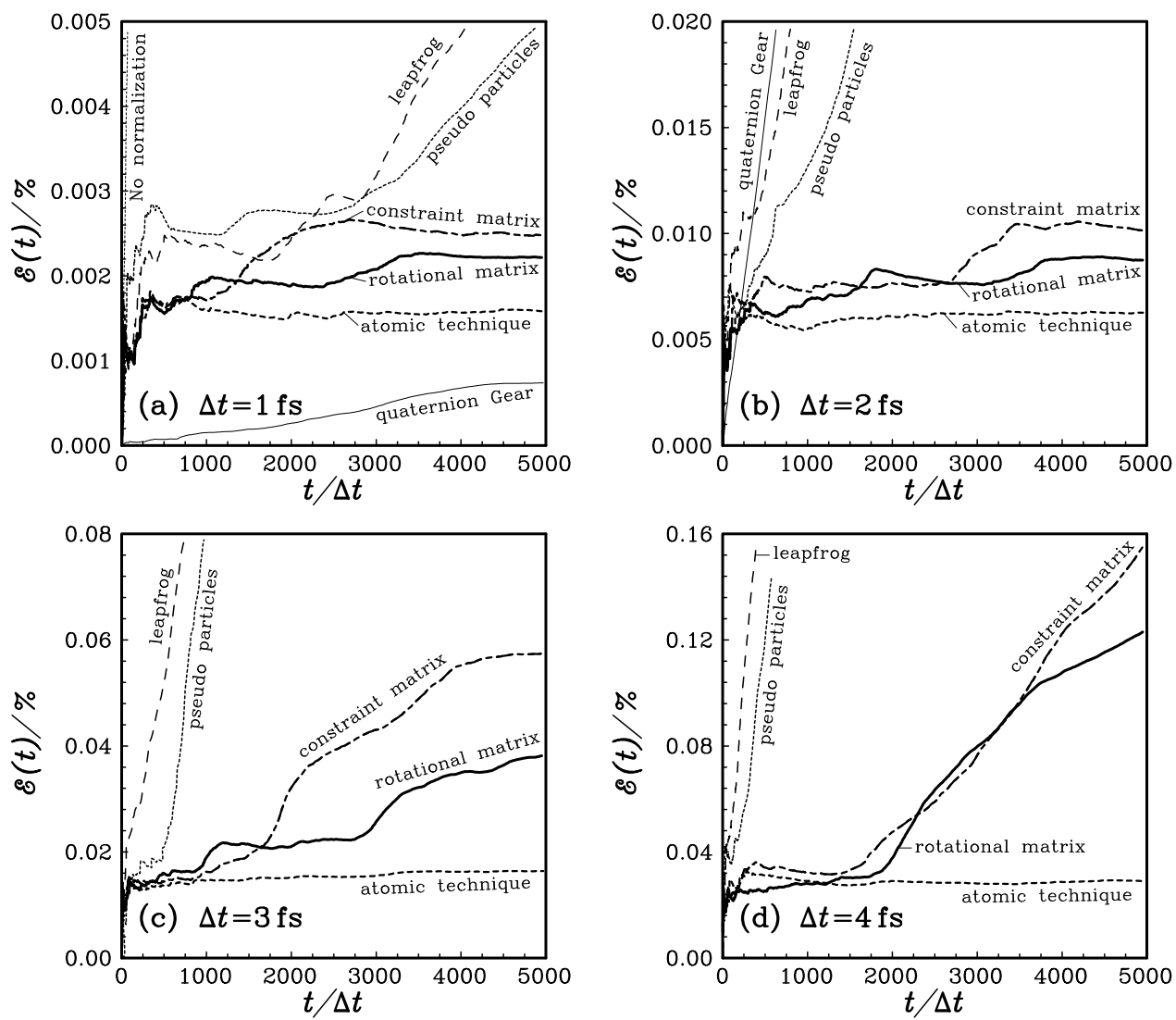
{\bf Fig.~1.} The relative total energy fluctuations as functions of the
length of the simulations on the TIP4P water, obtained within various
techniques at four fixed time steps, namely, 1 fs {\bf (a)}, 2 fs
{\bf (b)}, 3 fs {\bf (c)} and 4 fs {\bf (d)}.

\vspace{12pt}

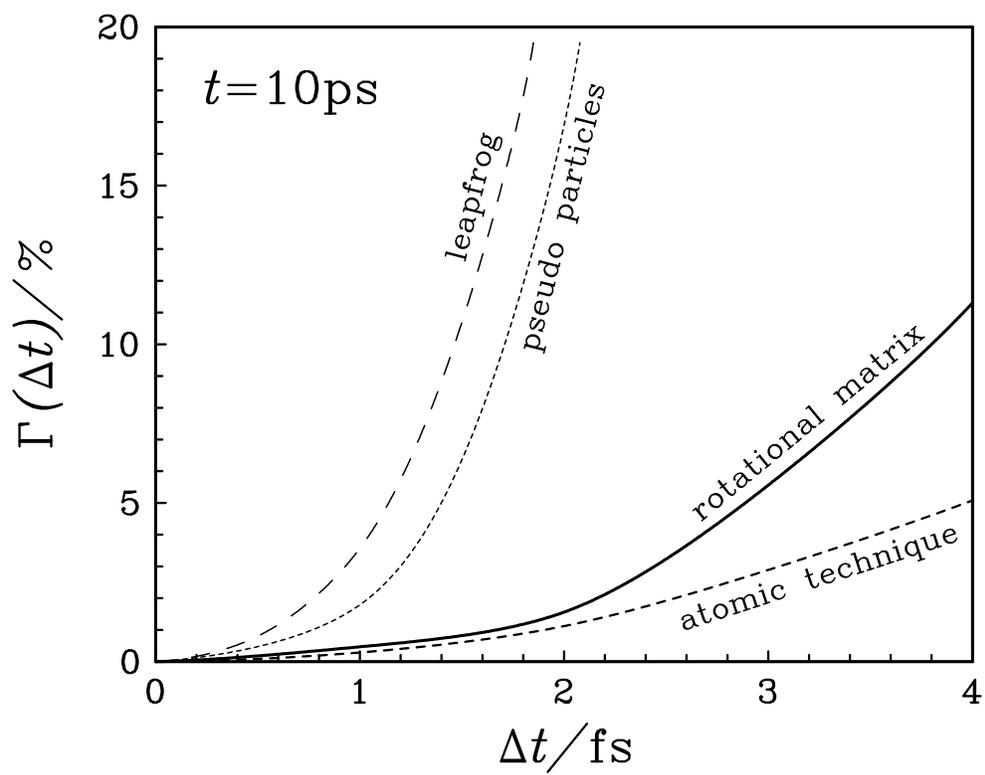
{\bf Fig.~2.} The ratios of the total energy and potential energy
fluctuations as dependent on the step size, observed for various
approaches in the simulations of the TIP4P water at the end of
10 ps runs.

\end{document}